\documentclass[cits]{PoS}
\usepackage{amsmath,bbm,slashed,dsfont}
\usepackage{wrapfig}

\usepackage{mciteplus}

\newcommand{\be}{\begin{equation}}
\newcommand{\ee}{\end{equation}}
\newcommand{\Z}{\mathcal{Z}}
\newcommand{\D}{\mathcal{D}}
\newcommand{\expv}[1]{\left \langle #1 \right \rangle}
\newcommand{\tr}{\textmd{tr}}
\renewcommand{\d}{\textmd{d}}

\renewcommand{\O}{\mathcal{O}}

\newcommand{\mv}{m_{ud}}
\newcommand{\ms}{{\widetilde{m}_{ud}}}

\newcommand{\mss}{{\widetilde{m}_{s}}}

\title{Chiral transition via the Banks-Casher relation}

\ShortTitle{Chiral transition via the Banks-Casher relation}

\author{\speaker{Gergely Endr\H{o}di}\\
        Institute for Theoretical Physics, Goethe University, Max-von-Laue-Strasse 1, 60438 Frankfurt am Main, Germany \\
        E-mail: \email{endrodi@th.physik.uni-frankfurt.de}}

\author{Lukas Gonglach\\
        Institute for Theoretical Physics, Goethe University, Max-von-Laue-Strasse 1, 60438 Frankfurt am Main, Germany \\
        E-mail: \email{gonglach@th.physik.uni-frankfurt.de}}

\abstract{
We investigate the properties of the finite-temperature QCD transition towards the chiral limit using staggered quarks. Starting from the 2+1-flavor physical point, the limit of massless quarks is approached along two different trajectories in the Columbia-plot. Unlike in previous approaches, the chiral condensate is determined via the Banks-Casher relation. The first results of our finite size scaling analysis are presented. 
}

\FullConference{The 36th Annual International Symposium on Lattice Field Theory - LATTICE2018\\
		22-28 July, 2018\\
		Michigan State University, East Lansing, Michigan, USA.}

\begin{document}

\section{Introduction}

The Columbia-plot~\cite{Brown:1990ev} depicts the nature of the finite-temperature chiral transition of QCD 
as a function of the quark masses. It is a fundamental diagram that reflects the 
properties of the chiral symmetry breaking mechanism of the strong interactions~\cite{Stephanov:2004wx} and is therefore interesting 
from the theoretical point of view.
Furthemore, regions with real phase transitions in 
the Columbia-plot are also related to critical behavior in the QCD phase diagram, 
for example, in the presence of baryon chemical potentials $\mu$. For purely imaginary $\mu$,
this critical region involves the Roberge-Weiss phase transition, while for real chemical 
potentials the conjectured critical endpoint of QCD, sought for extensively in current 
heavy-ion collision experiments. For recent reviews on the Columbia-plot and its applications,
see Refs.~\cite{Szabo:2014iqa,*Meyer:2015wax,*Bazavov:2015rfa,*Ding:2017giu,deForcrand:2017cgb}.

\begin{figure}[b]
 \centering
 \vspace*{-.3cm}
 \mbox{
 \includegraphics[width=.47\textwidth]{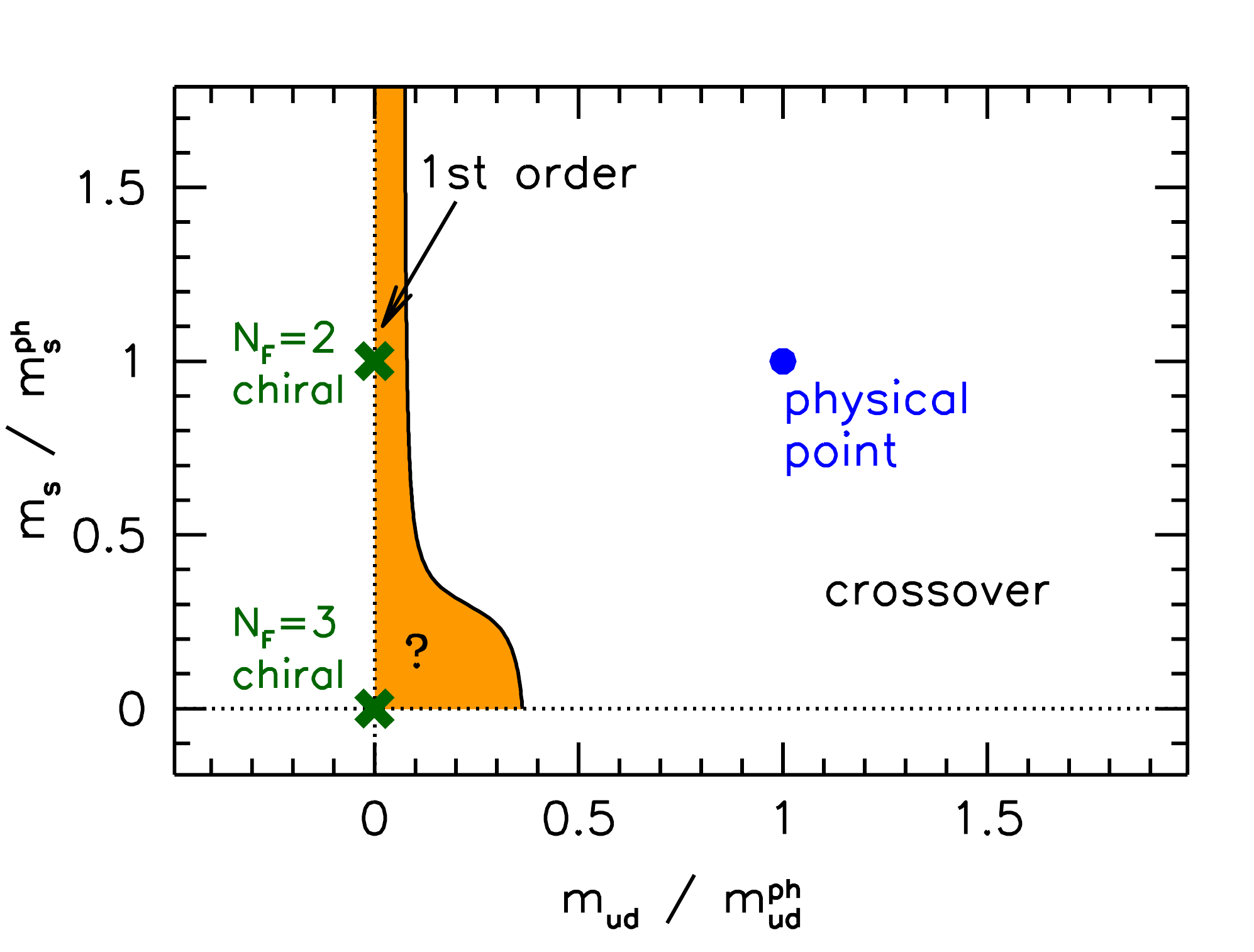}\quad
 \includegraphics[width=.47\textwidth]{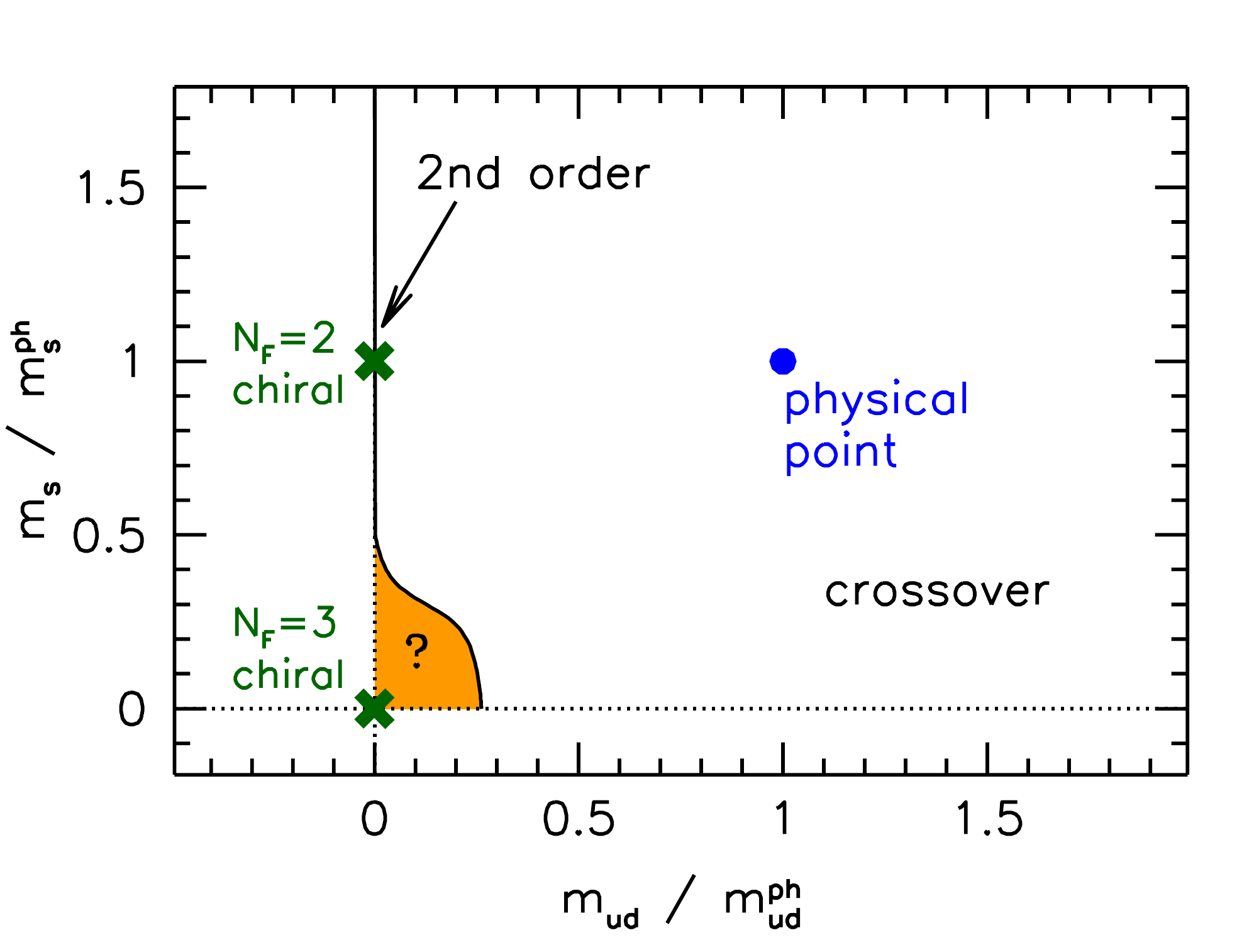}
 }
 \vspace*{-.2cm}
 \caption{\label{fig:1}
 Two possible scenarios for the lower left corner of the 
 Columbia-plot. Conjectured regions with a first-order phase transition (orange) are separated from the 
 crossover domain (white), including the physical point (blue dot) by second-order transition lines (solid black). Of interest is the nature of the transition in the $N_f=2$ and $N_f=3$
 chiral points (green crosses). }
\end{figure}

Assuming degenerate up and down quarks, the relevant parameters characterizing the Columbia-plot
are $m_{ud}$ and $m_s$. Here we concentrate on the lower left corner of the plot (see Fig.~\ref{fig:1}), where 
all quark masses are smaller than their
physical values $m_{ud}^{\rm ph}$ and $m_s^{\rm ph}$. 
At the physical point the transition is known to be an analytic crossover~\cite{Aoki:2006we,*Bhattacharya:2014ara}.
According to a low-energy effective model of QCD~\cite{Pisarski:1983ms}, in the $N_f=3$ chiral limit (the origin
of the Columbia-plot), the transition ought to be first order, extending also to nonzero values 
of the quark masses in both directions. Increasing the quark masses or reducing the number of 
massless flavors is expected to weaken the transition. Whether or not 
the first-order region 
includes the $N_f=2$ chiral point is of particular relevance, as this question 
is related to the effective restoration of the $\mathrm{U}(1)_{A}$ symmetry at high temperatures~\cite{Szabo:2014iqa,*Meyer:2015wax,*Bazavov:2015rfa,*Ding:2017giu}. 

The $N_f=2$ and $N_f=3$ chiral limits have been studied using different approaches 
in the literature. On the one hand, results obtained using coarse lattices and unimproved actions 
point towards the scenario depicted in the left panel of Fig.~\ref{fig:1}.
In particular, a second-order critical point at $m_{ud}>0$ has been 
identified using staggered quarks, for example in Refs.~\cite{Bonati:2014kpa,Cuteri:2017gci}.
However, lattice discretization effects appear to be dramatic, with a significant reduction 
of the first-order region as the lattice is made finer and no firm conclusion about what 
happens in the continuum limit.
Similar tendencies are visible~\cite{deForcrand:2017cgb} 
 in the three-flavor and four-flavor (not represented 
in the Columbia-plot) theories (see, e.g., Refs.~\cite{Cuteri:2017gci,Karsch:2001nf,*Varnhorst:2015lea,*Jin:2017jjp})
using both Wilson and unimproved staggered quarks.
On the other hand, simulations employing improved lattice actions observe no direct sign of a critical point, but only a strengthening
of the crossover transition as the quarks are made lighter (see, e.g.\ Refs.~\cite{Endrodi:2007gc,Burger:2011zc,Ding:2018auz,Brandt:2016daq}). Comparing the data to 
the critical behavior around a second-order point with an expected universality class 
(potentially: $\mathrm{Z}(2)$, $\mathrm{O}(2)$, $\mathrm{O}(4)$ or $\mathrm{U}(2)\otimes\mathrm{U}(2)/\mathrm{U}(2)$) in principle enables one to 
distinguish between the two scenarios of Fig.~\ref{fig:1}. However, a reliable 
determination of the critical mass turns out to be very difficult in practice. 

\section{A new method for approaching the chiral limit around the transition}

In the present talk we attempt to extrapolate the results of the simulations at nonzero 
quark masses to the chiral limit in order to learn about the nature of the phase transition
directly at the points of interest. We employ $2+1$ flavors of stout-improved rooted 
staggered quarks with masses around the physical point. 
Using two sets of ensembles, we approach the $N_f=2$ chiral point using simulations 
at fixed $m_s=m_s^{\rm ph}$ and the $N_f=3$ chiral point keeping the ratio
$m_s/m_{ud}=m_s^{\rm ph}/m_{ud}^{\rm ph}=28.15$ constant.
The employed quark masses lie in the range $0.25\cdot m_f^{\rm ph}\le m_f\le1.5\cdot m_f^{\rm ph}$ 
for both quark flavors $f=ud,s$, corresponding to approximate pion masses 
$68\textmd{ MeV}<m_\pi<170 \textmd{ MeV}$. 
The simulations are performed 
using $16^3\times6$ and $24^3\times6$ lattices in order to carry out a preliminary 
finite size scaling analysis. A temperature scan is carried out for each of the 
simulation points,
by using $8-10$ different temperatures (i.e.\ different 
inverse gauge couplings $\beta$). For each of these scans, the quark masses are
tuned along the lines 
of constant physics $m_{f}(\beta)$~\cite{Borsanyi:2010cj} and the lattice scale $a(\beta)$ 
is set at the physical point.

The order parameter (or, for massive quarks, the approximate order parameter) 
for chiral symmetry breaking is the light quark condensate
$\expv{\bar\psi\psi}= T/V\cdot \partial \log\Z / \partial m_{ud}$,
where $\Z$ is the partition function of the system, $T$ the temperature and $V$ the 
spatial volume. 
Up to an overall normalization factor, the condensate 
has the path integral representation
\be
\expv{\bar\psi\psi(\mv)}_{\ms,\mss}
= \frac{T}{2V} \int \D U e^{-S_g[U]} \,[\det{} ( \slashed{D}+\ms) ]^{1/2}\, [\det{} ( \slashed{D}+\mss) ]^{1/4} \,\tr  (\slashed{D}+\mv)^{-1} \,,
\label{eq:pbpdef}
\ee
where we used rooting and highlighted two sources of dependence 
on the quark masses: via the operator (valence quark mass $m_{ud}$) and via the 
quark determinants (sea quark masses $\widetilde{m}_{ud}$ and
$\widetilde{m}_s$). 
Our strategy for the extrapolation of $\expv{\bar\psi\psi}$ to the chiral points 
involves two distinct methods: 
the Banks-Casher relation and leading-order reweighting. The former will be used to extrapolate 
the valence quark mass to zero, while the latter to approach the chiral point in 
the sea quark masses. 
Both of these methods rely on well-known theoretical concepts and have been widely employed 
in the literature, particularly at low temperatures. Here we use them in a combined manner, 
in the context of the finite temperature QCD transition for the first time. 
This extrapolation strategy is inspired by a similar approach for 
QCD with isospin chemical potentials, where it proved highy 
beneficial for the determination of the order parameter for pion condensation~\cite{Brandt:2017oyy}.

The reweighting of an arbitrary observable $A$ towards the $N_f=2$ chiral point amounts to
\be
\expv{A}_{0,\mss} = \frac{\expv{A \,W}_{\ms,\mss}}{\expv{W}_{\ms,\mss}},\quad\;\;\;\;\;
W\equiv\frac{[\det{} ( \slashed{D}) ]^{1/2}}{[\det{} ( \slashed{D}+\ms) ]^{1/2}}
= \exp\left[ -\frac{V}{T}\ms \cdot \bar\psi\psi(\ms) +\O(\ms^4) \right]\,,
\label{eq:rewfac}
\ee
where we expanded the reweighting factor in $\ms$, resulting, to leading order, 
in the exponential of the quark condensate at finite mass. The latter is evaluated 
using noisy estimators. This reweighting will be performed
below for the spectral density.
For approaching the $N_f=3$ chiral point, an additional reweighting in $\mss$ could 
in principle also be carried out. However we found that the latter results in large 
fluctuations due to the high strange quark mass and was therefore not performed.

\begin{figure}[b]
 \centering
 \vspace*{-.3cm}
 \mbox{
 \includegraphics[width=.48\textwidth]{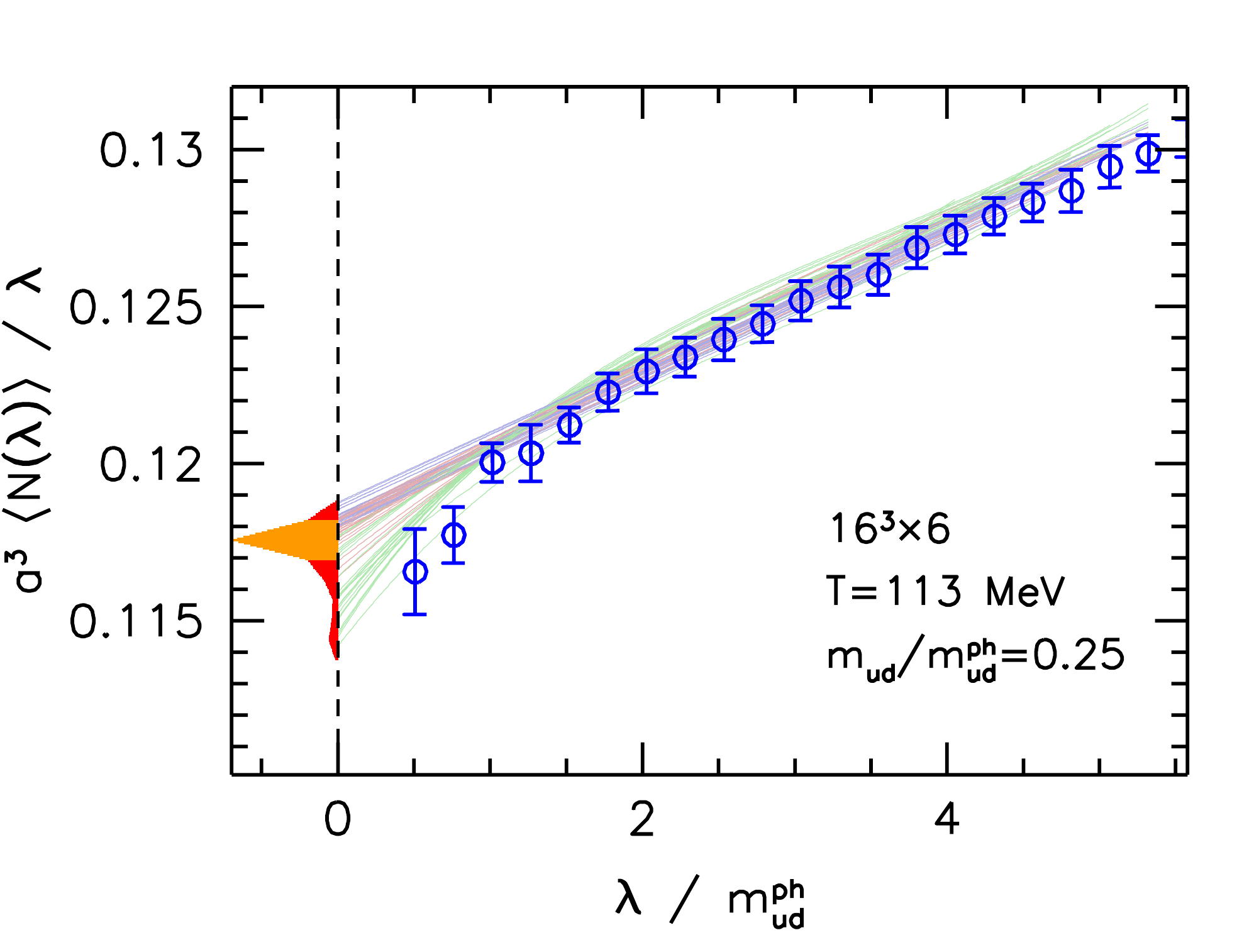}\quad
 \includegraphics[width=.48\textwidth]{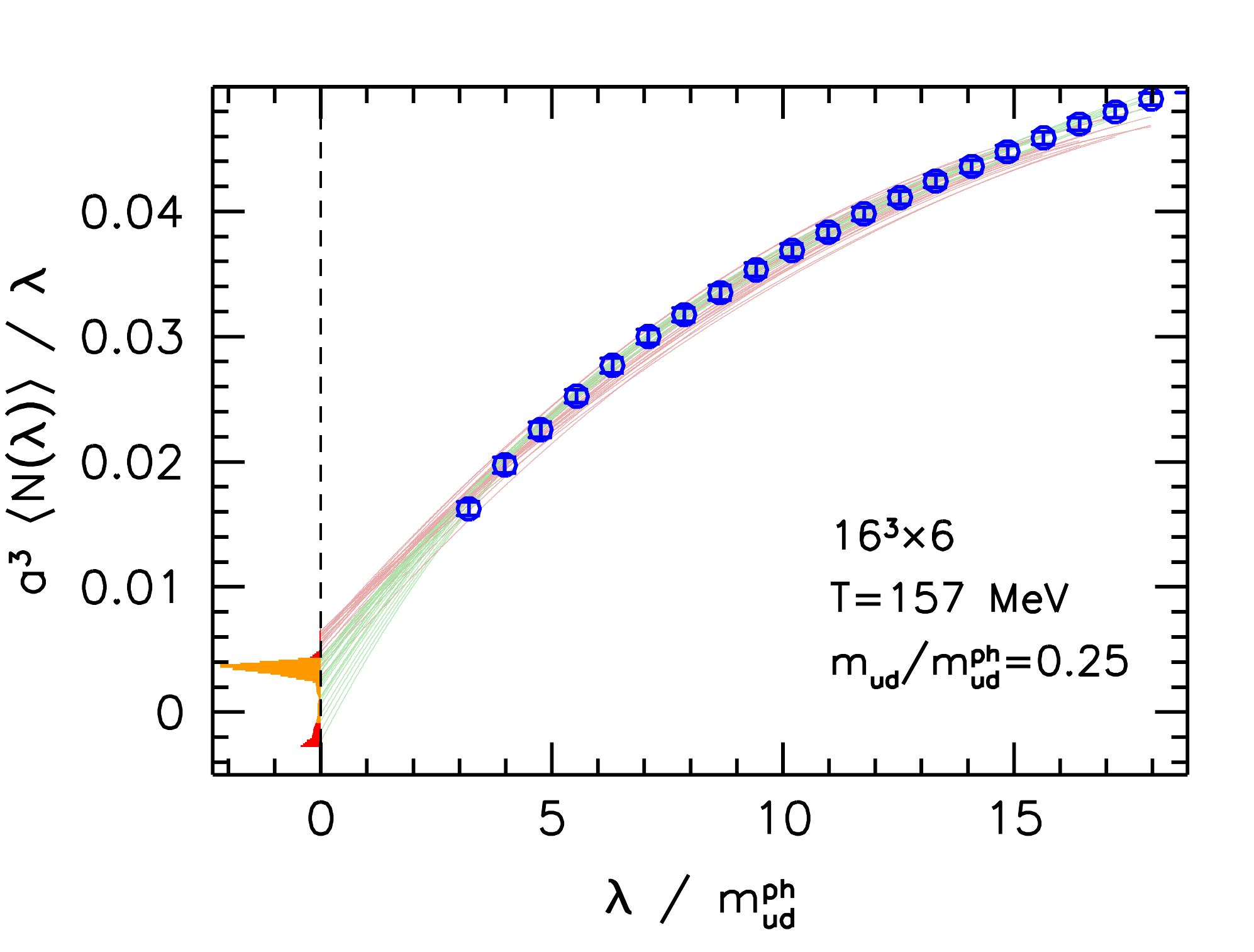}
 }
 \vspace*{-.4cm}
 \caption{\label{fig:3}
 The integrated spectral density well below (left panel) and around (right panel) the 
 transition temperature. The colored curves indicate polynomial fits
of the data: linear (blue, only in the left panel), quadratic (red) and cubic (green).
The red vertical histograms mark the distribution of $\rho(0)$ as
obtained from the fits. The orange part of the histogram ($68\%$ around the median) is
our estimate of the statistical plus systematical error of the
fits.}
\end{figure}

The Banks-Casher relation~\cite{Banks:1979yr} connects the operator appearing in the path integral~\eqref{eq:pbpdef} to the spectral density $\rho$ of the Dirac operator $\slashed{D}$ 
around the origin. Approaching consecutively the thermodynamic and the chiral limits, 
it can be succintly written as
\be
\frac{T}{V}\expv{\tr(\slashed{D}+\mv)^{-1}} = \frac{T}{V} \bigg\langle\sum_k \frac{\mv}{\lambda_k^2+\mv^2}\bigg\rangle
\xrightarrow{V\to\infty} \int_{-\infty}^\infty \!\d\lambda\expv{\rho(\lambda)} \frac{\mv}{\lambda^2+\mv^2}
\xrightarrow{\mv\to0} \pi\expv{\rho(0)}\,,
\label{eq:BC}
\ee
where the trace was expressed in the eigenbasis $\slashed{D}\,\psi_k=i\lambda_k\psi_k$ 
of the Dirac operator and 
$\rho$ is understood to include the normalization factor $T/V$.
We also define the integrated spectral density 
$\expv{N(\lambda)} = \int_{0}^{\lambda} \!\d\lambda' \expv{\rho(\lambda')}$,
which is better suited for an extrapolation $\lambda\to0$ as it is flatter around the origin than the spectral density
itself. 
We plot $\expv{N(\lambda)}$ normalized by the upper endpoint $\lambda$ of the integral 
in Fig.~\ref{fig:3} for temperatures well below and just around the transition 
region, as obtained on our $16^3\times 6$ lattices at $m_s=m_s^{\rm ph}$ and 
$m_{ud}=0.25\cdot m_{ud}^{\rm ph}$. 

Note that -- as~\eqref{eq:BC} highlights -- an infinite-volume 
extrapolation is necessary to relate $\expv{\rho(0)}$ to the chiral condensate. 
Indeed, in any finite volume the spectral 
density vanishes at the origin, since the smallest eigenvalues are of $\O(1/V)$. 
This is tantamount to the vanishing of the $m=0$ limit of the condensate -- i.e., 
to the absence of spontaneous symmetry breaking -- in finite volumes. 
Consequently, the determination of $\expv{\rho(0)}$ in finite volumes involves extrapolating the spectral density to $\lambda=0$. This is performed by means of polynomial 
fits of the data, which are combined to give a weighted histogram for $\expv{\rho(0)}$, also visualized in Fig.~\ref{fig:3}. 
We note that this $\O(1/V)$ extrapolation might bias the volume-dependence of 
the order parameter, and of its slope, which will be determined below.
To avoid this extrapolation, the volume scaling analysis 
could be performed already at the level of the spectral density at nonzero $\lambda$.
This will be investigated in the future.

\begin{wrapfigure}{r}{7.1cm}
 \centering
 \vspace*{-.3cm}
 \includegraphics[width=7.2cm]{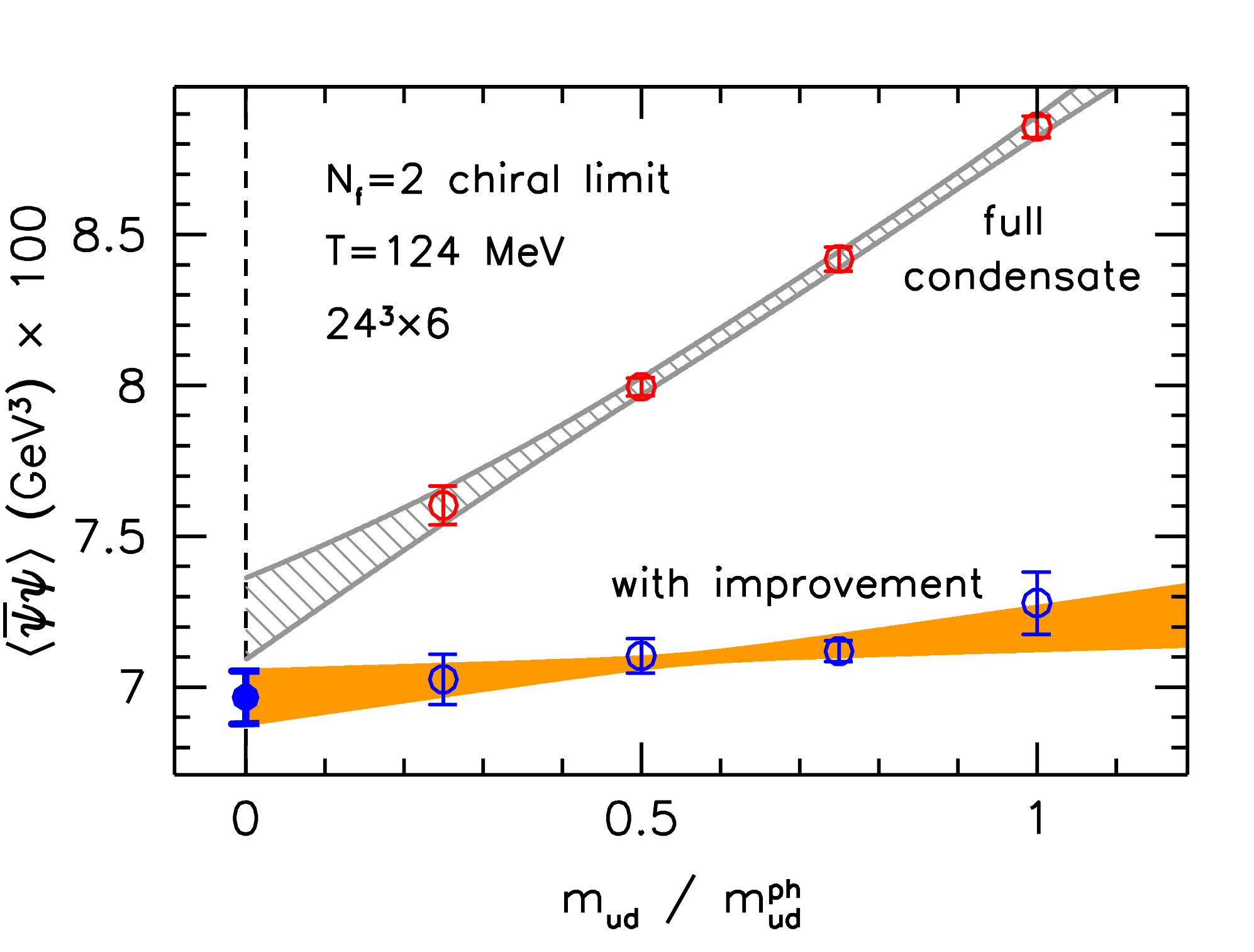}
 \caption{\label{fig:4}
 Residual dependence on the sea quark mass using our improvement scheme (blue points), 
 extrapolated to zero (yellow band). A comparison to the unimproved condensate (red points) and its extrapolation (gray dashed band) is also shown.
 }
 \vspace*{-.0cm}
\end{wrapfigure}

Having determined the reweighted $\expv{\rho(0)}_{0,\mss}$ in this manner, we can 
compare the results obtained on ensembles with different sea quark masses. A residual 
dependence on the quark mass still 
remains due to the reweighting being only performed 
to leading order and due to lattice artefacts. 
One example for this dependence is shown 
in Fig.~\ref{fig:4}, including a final extrapolation of the results to $m_{ud}=0$. 
For comparison, the results for the unimproved quark condensate (determined using 
noisy 
estimators according to Eq.~\eqref{eq:pbpdef}) is also included. Clearly, 
our improvement scheme 
performs 
much better and facilitates a 
controlled extrapolation. The extrapolation of the unimproved 
data is found to slightly 
overestimate the chiral limit. A similar procedure is performed for 
approaching the $N_f=3$ 
chiral point. Although still much milder
than for a naive extrapolation, in this case the 
residual mass-dependence is stronger due to the higher strange quark mass and the 
exclusion of the reweighting factors for the strange quark (see above).
We remark that while the chiral order parameter should be non-negative, the result of the 
above described double extrapolation may be negative. In such cases the positive part of the 
error bar is taken as the uncertainty of the result.

\section{Finite size scaling of the chiral order parameter}

\begin{figure}[t]
 \centering
 \vspace*{-.0cm}
 \mbox{
 \includegraphics[width=.48\textwidth]{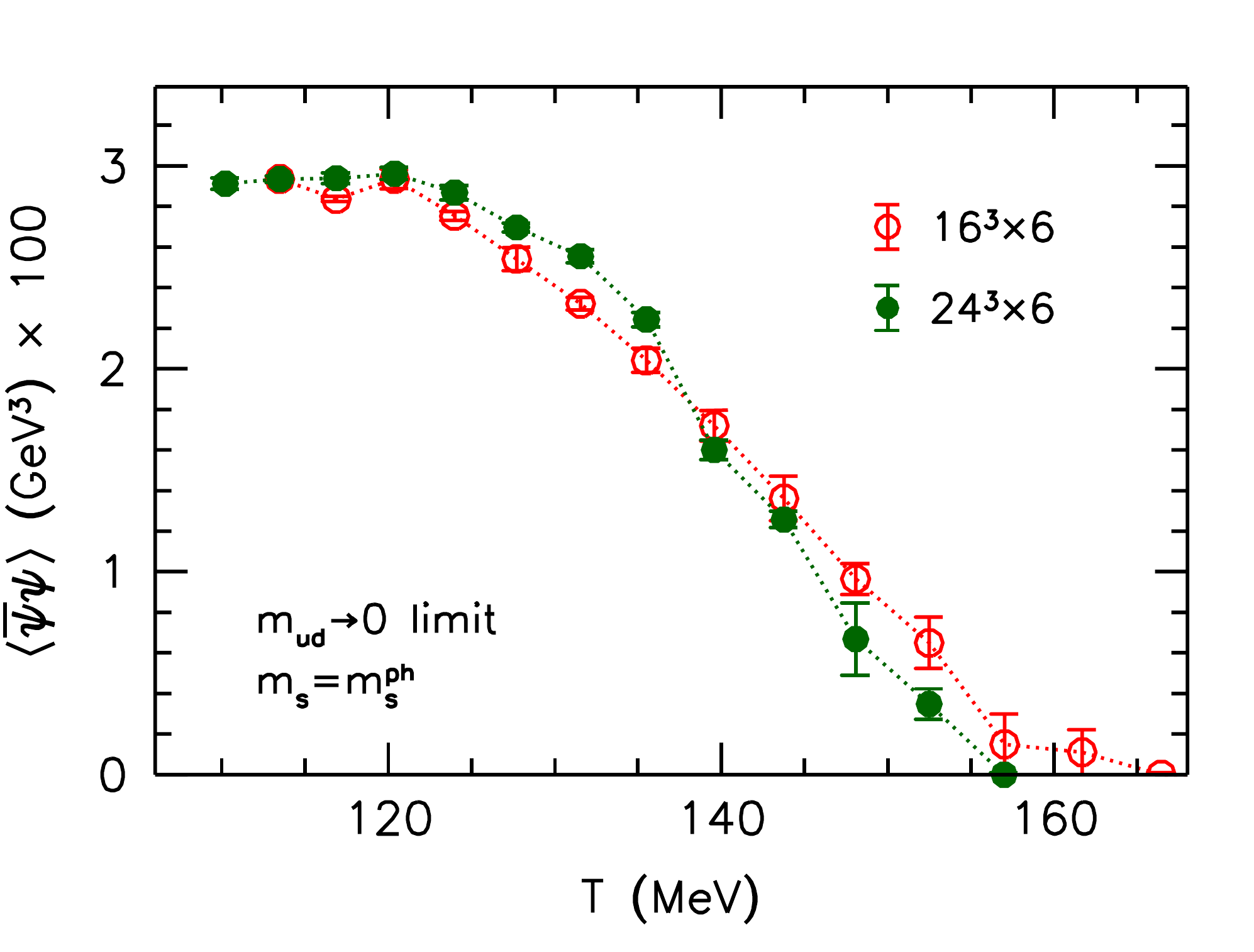}\quad
 \includegraphics[width=.48\textwidth]{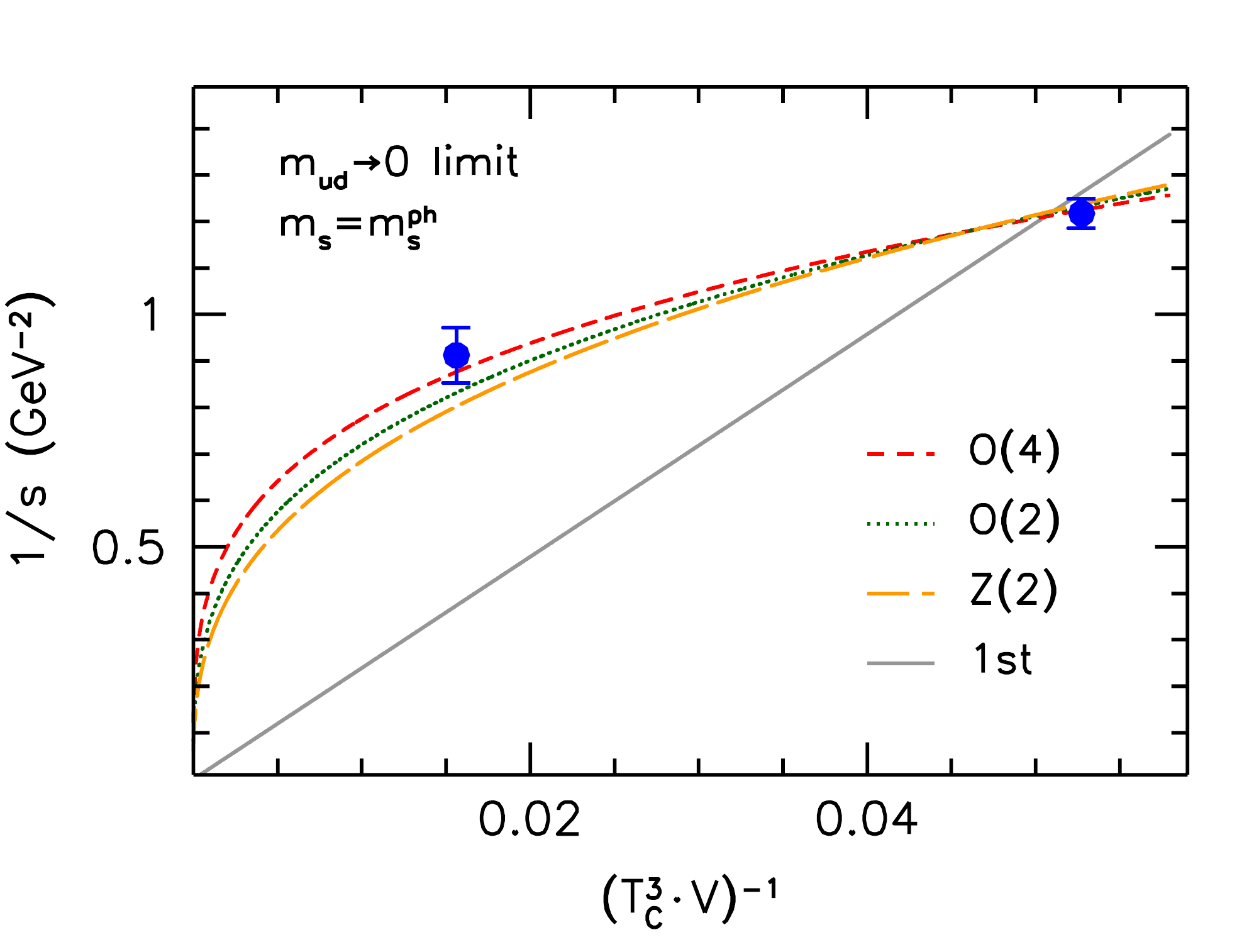}
 }
 \vspace*{-.3cm}
 \caption{\label{fig:5}
 Left panel: the chiral order parameter, after an extrapolation to the $N_f=2$ chiral point 
 for two different volumes.
 Right panel: finite size scaling of the inverse slope of the order parameter at the critical 
 temperature, compared to the behavior for a first-order transition and 
 for second-order phase transitions
 for different universality classes.
 }
\end{figure}

Our final results for the extrapolated order parameter are shown in the left panel of Fig.~\ref{fig:5}
for the $N_f=2$ chiral limit. The currently available two volumes, 
$16^3\times6$ and $24^3\times6$, are compared in the interesting region $110\textmd{ MeV}<T<160\textmd{ MeV}$. The data exhibits a plateau for 
low temperatures, which corresponds to $\expv{\bar\psi\psi}^{N_f=2}\approx(310 \textmd{ MeV})^3$, 
in the ballpark of typical values for the light quark condensate for two chiral 
and one massive flavors.
Notice that unlike the quark condensate at nonzero mass, which always involves explicit 
symmetry breaking terms, our order parameter 
approaches zero in the thermodynamic limit 
for high $T$ where the spontaneously broken chiral symmetry is restored. 
We also emphasize that the chiral order parameter contains no additive divergences,
so that unlike for the massive condensate, no additive renormalization is required.

Comparing the two different volumes reveals a sharpening of the order parameter near 
the transition, hinting at the presence of a singularity in the thermodynamic limit.
Indeed, for a real phase transition the order parameter drops to zero at $T_c$ with infinite 
slope. For finite volumes this behavior is smoothed out, with a 
transition temperature that approaches $T_c$ and a slope that diverges as $V\to\infty$. 
The results suggest the ballpark value $T_c^{N_f=2}\approx (135  - 150) \textmd{ MeV}$, which lies close to the recent determination based on 
a scaling analysis of susceptibilities at nonzero quark mass~\cite{Ding:2018auz}.
(We stress that to express the temperature and the condensate in dimensionful 
units, the lattice scale at the physical point was used.)

To be more quantitative regarding the sharpening of the order parameter, we determine the negative slope $s\equiv-\partial\expv{\bar\psi\psi}/\partial T|_{T=T_c}$ at the critical temperature by fitting the data in the transition region 
using a linear function. The way that $s$ diverges for $V\to\infty$ determines
the order of the phase transition. For a first-order transition the scaling is linear 
in the volume: $s\propto V$, while for a second-order transition it is dictated by a combination 
of the critical exponents $\beta$ and $\nu$: $s\propto V^{(1-\beta)/(3\nu)}$.
(Note that while here we only investigate the 
slope, the complete dependence of the order parameter on its variables around $T_c$
is fixed according to critical scaling.)
We plot the results for the inverse slope $1/s$ against the inverse volume 
$1/V$ in the right panel of Fig.~\ref{fig:5}. 
The data is fitted according the relevant critical 
exponents~\cite{0305-4470-28-22-007,*Kanaya:1994qe,*Campostrini:2000iw}. Notice that 
the only free parameter for these fits is the normalization of the curves. 
The results appear to prefer a second-order phase transition 
with the $\mathrm{O}(4)$ universality class, although the other considered universality 
classes also lie close to the data. 
Repeating the same analysis with an extrapolation to the $N_f=3$ chiral point 
(using ensembles with $m_s/m_{ud}$ fixed),
we observe a similar trend, with a slope that increases more strongly towards the thermodynamic 
limit as for the $N_f=2$ chiral point.

\section{Summary}

In this talk we presented a novel approach for the investigation of the chiral phase transition 
in QCD. The order parameter of the transition is determined via an extrapolation towards the 
chiral limit using the Banks-Casher relation and leading-order reweighting. Our preliminary finite size 
scaling analysis, based on two lattice volumes, shows a significant enhancement of the slope of the 
order parameter as the volume grows. 
The comparison of the results with the expected critical scaling suggests that 
the transition is of second order, although no
firm conclusions can be made: at least one larger volume will be necessary in order to 
perform the scaling fits in a controlled and reliable manner. 
The methods developed in the present contribution should be employed for finer lattices 
in order to enable a continuum extrapolation, and also for other fermion discretizations
in order to test universality towards the continuum limit.\\

\noindent
{\bf Acknowledgments}\; 
This research was funded by the DFG (Emmy Noether Programme EN 1064/2-1).
The majority of the simulations was performed on
the FUCHS cluster at the
Center for Scientific Computing of the Goethe University of Frankfurt.
The authors are grateful for enlightening discussions with Bastian Brandt, Francesca Cuteri,
Philippe de Forcrand, S\'andor Katz, Frithjof Karsch, Tam\'as Kov\'acs, Swagato Mukherjee, D\'aniel N\'ogr\'adi and 
Alessandro Sciarra.

\bibliographystyle{JHEP_mcite}
\bibliography{columbia_pos}

\end{document}